\documentclass[12pt]{article}

\usepackage{latexsym, amssymb, amsfonts,amsmath}
\usepackage{graphicx}
\usepackage{subcaption}
\usepackage{enumitem}

\newcommand{\negr}[1]{\boldsymbol{#1}}

\usepackage{amssymb}
\usepackage{hyperref} 

\usepackage{adjustbox} 
\usepackage[english]{babel} 
\usepackage{verbatim}
\usepackage{latexsym}
\usepackage{graphics}
\usepackage{epsfig} 
\usepackage{graphicx}
\usepackage{fancybox} 
\usepackage{ifsym} 
\usepackage{multirow,bigdelim} 
\usepackage[cmtip,arrow]{xy} 
\usepackage{pb-diagram,pb-xy} 
\usepackage{comment} 
\usepackage{cjhebrew}  

\def\modulo{\;\mbox{mod}\,}

\def\rr{\noindent - }

{\list{}{\leftmargin=0.07in\rightmargin=0.07in}\item[]}%
{\endlist}

\title{Randomization and Fair Judgment \\ in Law and Science}

\author{
	Julio Michael Stern\footnote{Institute of Mathematics and Statistics of the University of Sao Paulo \texttt{jstern@ime.usp.br}} , \\ 
	Marcos Antonio Simplicio\footnote{Polytechnic School of the University of Sao Paulo \texttt{mjunior@larc.usp.br}} ,  \\  
	Marcos Vinicius M. Silva\footnote{Polytechnic School of the University of Sao Paulo \texttt{mvsilva@larc.usp.br}} , \\ 
	Roberto A. Castellanos Pfeiffer\footnote{Law School of the University of Sao Paulo, \texttt{roberto.pfeiffer@usp.br}} \\ }

    \date{\small 
 p.399-418 in
 Jos\'{e} Ac\'{a}cio de Barros, D\'{e}cio Krause, eds. 
 \textit{A True Polymath: A Tribute to Francisco Ant\^{o}nio D\'{o}ria}.  Rickmansworth, UK: College Publications. ISBN: 978-1-84890-351-7}

\usepackage[paperheight=297mm, paperwidth=210mm,  
layoutheight=200mm, layoutwidth=120mm,
layoutvoffset=41.9mm, layouthoffset=45mm, 
centering,
margin=0pt, includeheadfoot,
footskip=5mm,
]{geometry}

\usepackage{microtype}

\begin{document} 		  
	
	\maketitle

	\begin{abstract}  
		Randomization procedures are used in legal and statistical applications, aiming to shield important decisions from spurious influences.     
		This article gives an intuitive introduction to randomization and examines some intended consequences of its use related to truthful statistical inference and fair legal judgment.   
		This article also presents an open-code Java implementation for a cryptographically secure, statistically reliable, transparent, traceable, and fully auditable randomization tool.  
		
	\end{abstract} 
	
	\pagebreak 

	\begin{flushright}

		{$\kappa \lambda \eta \rho \omega \   \nu \upsilon \nu \    
			\pi \epsilon \pi \alpha \lambda \alpha \sigma \theta \epsilon \  
			\delta \iota \alpha \mu \pi \epsilon \rho \epsilon \varsigma \ 
			o \varsigma \  \kappa \epsilon \  
			\lambda \alpha \chi \eta \sigma \iota \nu \! :$ } \\ 
		\mbox{} \vspace{-4mm} \mbox{} \\ 
		\textit{Let the lot be shaken for all of you, \\ 
			and see who is chosen.}     Iliad, VII, 171.  \\   
		\mbox{} \vspace{0mm} \mbox{}  \\  
		
		\<mid:yAniyM  ya+s:b*iyt hag*worAl UbeyN `:a.sUmiyM yap:rid> \\ 	\mbox{} \vspace{-4mm} \mbox{} \\ 
		\textit{Casting the dice puts judgment quarrels to rest and keeps \\ 
			distinct essential powers duly separated.}    \ Proverbs 18:18.  \\   
	\end{flushright} 
	
	\mbox{} \vspace{0mm} \mbox{}

	\section{Introduction}  
	
	Francisco Antonio D\'{o}ria has had a consistent interest in randomness and chaos and, together with his collaborators, has investigated fundamental aspects of such phenomena.
	This article is our contribution to the Festschrift celebrating Doria's 75th birthday.  
	    
	This article analyses some pragmatical aspects of applying randomization in empirical science and law, considers some philosophical implications or premises justifying or motivating these applications, and offers some tools that promote good randomization practices.    
	\textit{The Cardsharps} (1594) marks the beginning of the independent career of the great Italian master Michelangelo Merisi da Caravaggio (1571-1610).    
	This painting displays a wealthy but innocent looking boy playing cards with his opponent, a cardsharp, that cheats in two ways: On the one hand, the cardsharp hides in his belt spurious cards that he intends to use in illegitimate ways; on the other hand, a sinister looking and strategically positioned accomplice gives him access to privileged and undue information.     
	Finally, the cardsharp carries a dagger, hinting at the dangers lurking in this environment of misrepresentation and deception.  
	
	\begin{figure}[tb] 
		\centering   
		\includegraphics[height=80mm, 
		trim= 0mm 0mm 0mm 0mm, clip]{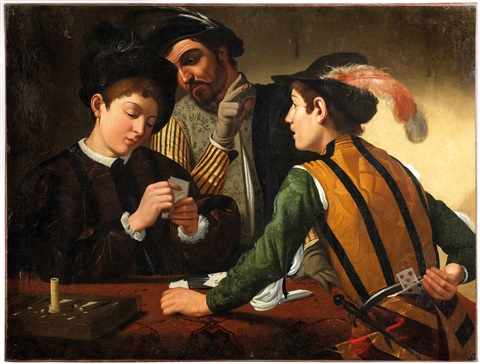} 
		
		\caption[]{\textit{The Cardsharps} (1594), by Michelangelo Caravaggio}    
		
	\end{figure}
	
	Caravaggio gives a beautiful depiction of some themes discussed in this article. 
	First, the social importance of activities involving randomization, that is, the random setting of some variable of interest, like the drawing of dice or, in this painting, the distribution of playing cards.    
	Second, it suggests the question -- Why to randomize? that is -- Why should a rational agent abdicate the opportunity of making a deterministic choice introducing, on purpose, a random component in making a decision?  
	If so -- What is the role played by randomization?  
	Finally -- How to randomize? that is -- What dangers could jeopardize a randomization process? and, if necessary -- How to shield or immunize the process against these dangers?  
	
	In order to answer these questions, we have to pay attention to some topics in Statistics, Computer Science and Cryptography; in addition, we have to examine some details concerning the design of empirical trials or the operation of legal systems.  
	In this article, we investigate each one of the questions just raised, looking for an intuitive understanding of the role(s) played by randomization. 
	
	In the final sections of the paper, we present an easy to use, open-code, traceable, auditable, secure, and statistically sound randomization toll that is ready for use in empirical trials and legal applications. 
	This kind of secure randomization tool can prevent the possibility of misrepresentation and deception, as depicted in the painting by Caravaggio. 
	Moreover, even in situations where no misdeeds actually occur, the use of such a tool can be beneficial by fostering public confidence in the soundness of important decisions, by strengthening the resilience of public institutions, and by favoring the peaceful resolution of conflicts.      
	
	\section{Social Importance of Randomization}

	Gambling and lotteries exchange billions of Dollars every day worldwide. 
	Hence, ensuring honesty and transparency in these activities should already be considered a meritory task. 
	However, since ancient times, sortition (i.e., selection by lottery) is used for many other purposes. 
	In the Iliad, one of the oldest texts of western culture (aprox. 1200BC), the Argonauts (crew of the ship Argo) selected a man to execute a dangerous task by sortition -- see this paper's first opening quotation.      
	In the same manner, modern societies often resort to sortition for drafting.  
	Figure 2 displays some photographs related to compulsory enlistment for service in the USA, namely, military drafting during the Civil (left) and the Vietnam (center) wars, and selection for jury duty (right).   
	
	In order to gain public trust, the sortitions for the Vietnam war were conducted in public view: Balls with calendar dates were placed in a transparent urn and some anniversary dates were then picked, giving the (un)lucky winners the opportunity to serve their country in the battlefield.   
	A post hoc statistical analysis of these drawings revealed a significant bias favoring latter days of the calendar, corresponding to the last balls placed inside the urn, an unexpected effect of an ill-conceived randomization process that generated misunderstanding, frustration and conflict.

	\begin{figure}[tb] 
		\centering   
		\includegraphics[height=36mm, 
		trim= 0mm 0mm 0mm 0mm, clip]{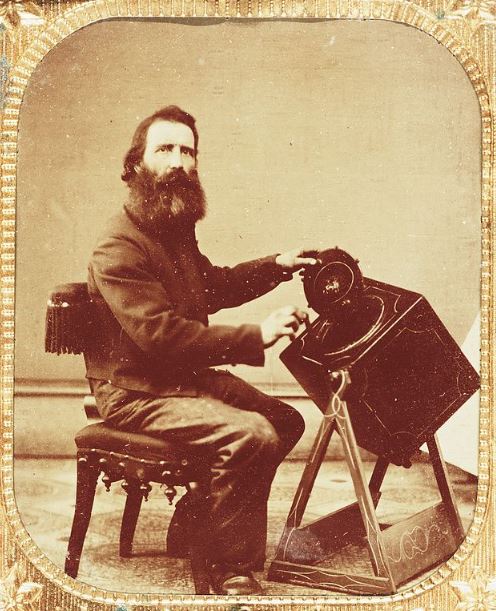} 
		\mbox{} \hspace{1mm} \mbox{}  
		\includegraphics[height=36mm,   
		trim= 0mm 0mm 0mm 0mm, clip]{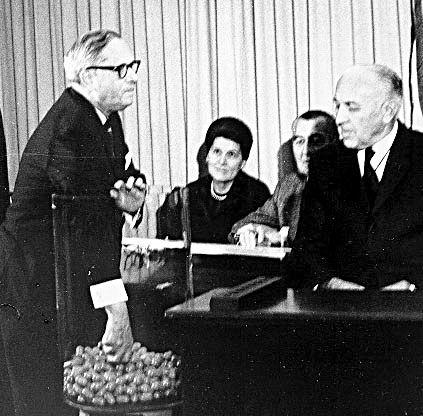} 
		\mbox{} \hspace{1mm} \mbox{}  
		\includegraphics[height=36mm,   
		trim= 0mm 0mm 0mm 0mm, clip]{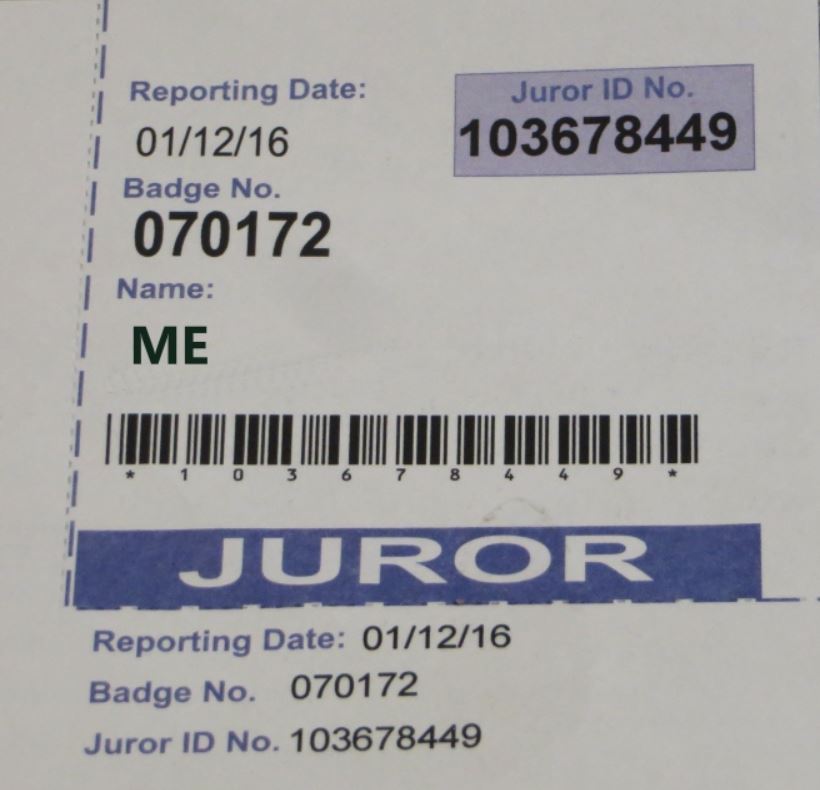} 
		
		\caption[]{Draft lotteries in war and peace}    
		
	\end{figure}
	
	Figure 2 (right) shows a letter calling a citizen for jury duty in the USA. 
	In this process, an eligible citizen was chosen at random by running a computer program. 
	Post hoc analyses of this randomization process revealed no significant bias or any other statistical anomaly. 
	Nevertheless, the code of these computer programs were never made public, making the randomization process opaque and non-verifiable, thus generating mistrust and resentment.

	Finally, in the world of science, good clinical trials are conducted by (double) blind and random attribution of patients to two or more distinct treatments. 
	The objective of such a trial is to find out if a new or alternative treatment is significantly better than the old or standard one, according to well-established statistical criteria. 
	In this situation, some frequently asked questions are: 
	Why should a patient's treatment be selected at random? 
	Why not give him or her the freedom to chose his or her proffered treatment?   
	Why hide from a patient information about his or her own treatment?

	\section{Why to Randomize?}  
	
	Imagine a clinical trial where patients are free to choose a treatment according to their own will. 
	Among patients, there will be rich and poor, people with different degrees of instruction, people with better or worst networks for support, etc. 
	Obviously, rich, well educated, and well connected patients will have better access to good information and advice and, therefore, will be prone to make better decisions. 
	Moreover, these same patients likely have better overall living conditions and, therefore, even with the same treatment, might have a better chance of recovery.
	Hence, this freedom of choice would automatically introduce \textit{confounding effects}:  
	After the trial is over, we would not be able to (completely) discern the beneficial and adverse consequences of distinct treatments from consequences of preexisting conditions. 
	
	Similar unwanted interference is generated by the \textit{placebo} and \textit{nocebo} confounding effects. 
	If a patient knows to be receiving either a new, experimental and possibly wonderful drug, or else an old and possibly not very effective drug, his or her moral may be, respectively, lifted or depressed.
	That, in turn, may affect his or her overall health and chance of recovery.     
	This is why, in a good clinical trial, treatment information is denied (blinding or censorship) to patients, and commonly also to their direct caretakers (double-blinding).

	There are in the medical literature plenty of examples of clinical trials that came to wrong conclusions in consequence of such confounding effects. 
	The best known antidotes against these confounding effects rely in some form of randomization. 
	The idea is to chose a patient's treatment based on a random variable that is independent of any potentially confounding variable.   
	In so doing, the random element in the choice of treatment has the effect of breaking causal links that should not interfere with the experiment, allowing the trial to adequately focus on the causal links of interest -- see Stern (2008) and Pearl (2009) for further details.

	Finally, let us consider the use of randomization in the legal system, like the selection of jurors or judge(s) for a given case. 
	Figure 3 displays two pictures from ancient Egypt. 
	On the left, a stone carving of approx. 2400BC shows two merchants using a two-pan balance to correctly measure amounts of goods for a fair commercial transaction. 
	On the right, the Hunefer papyrus, of approx. 1275BC, shows the scale used by Maat, the goddess of justice,  where the heart and the (de)merits of a man are measured. 
	It should be clear that these two scales are essentially distinct -- they belong to distinct contexts. 
	The figure at the center suggests the possibility of ``mixing'' these two essences: Perhaps Maat could make a more benevolent assessment in the scale of justice if she, or her priests, received goods of commercial valuable...
	There we have, once more, a confounding effect, characterized by spurious influences between powers belonging to essentially distinct systems: in this case, the economic system and the justice system.

	\begin{figure}[tb] 
		\centering   
		\includegraphics[height=30mm, 
		trim= 0mm 0mm 0mm 0mm, clip]{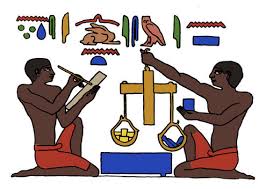} 
		\mbox{} \hspace{1mm} \mbox{}  
		\includegraphics[height=30mm, angle=0,  
		trim= 0mm 0mm 0mm 0mm, clip]{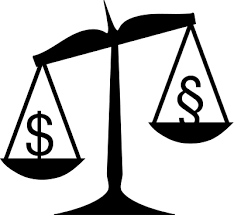} 
		\mbox{} \hspace{1mm} \mbox{}  
		\includegraphics[height=30mm, angle=0,  
		trim= 0mm 0mm 0mm 0mm, clip]{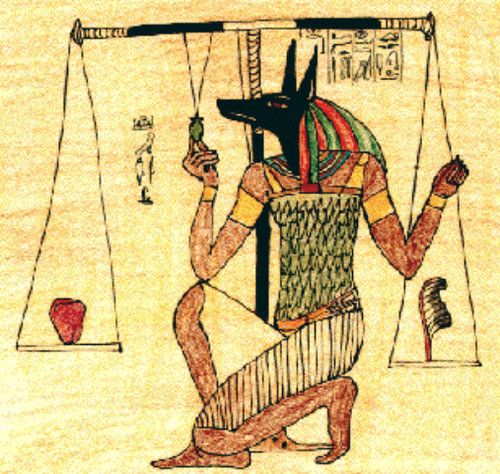} 
		
		\caption[]{Essentially different powers:  Economy and Justice}    
		
	\end{figure}

	How to avoid such confounding effects caused by spurious influences, fostering autonomous decisions in a strong and independent justice system? 
	Surprisingly, the Hebrew bible already offers very good advice at this respect, as stated in the second opening quotation of this paper.  
	Interestingly, the Hebrew root \<`.sm>, \textit{etzem}, whose literal meaning is bone, also generates words meaning essence (the etymological origin of the English word), strength, power and the modern Hebrew word for independence.

	Judges, even if perfectly honest, do not come to court as a blank slate, nor should they. 
	Every judge has his or her own history of decisions and opinions. 
	Hence, if the selection of judges could be influenced by the litigants or other interested parties, the richer, better informed, well connected, or otherwise more powerful parties would likely have an advantage in directing the case to a judge sympathetic to their arguments.   
	For this reason, in many modern democracies, the \textit{distribution} of a new judicial case must take the form of a random choice among the available judges or courts qualified to judge it.

	\section{How to Randomize?} 
	
	Previous sections discussed several applications of randomization and explanations of why to use it.   
	This section describes some desirable characteristics of such a randomization process, including: 
	\begin{description} 
		\item {Statistical honesty:} In a set of sortitions (random selections) in the system, the probability of any group of outcomes should be exactly as prescribed by the established rules. 
		
		\item{Cryptographic security:} The outcome of the sortition should be unpredictable; moreover, no external agent should be able to influence the randomization process, even if the agent knows in detail the randomization mechanism being used and has state of the art knowledge of all relevant technologies involved.   
		
		\item{Transparency:} All relevant information about the sortition process must be of public knowledge, including any pertinent detail about the randomization mechanisms being used. 
		
		\item{Auditability:} All relevant occurrences of an actual randomization process must be traceable and auditable.  
		Furthermore, it should not be possible to conceal any improper use of the randomization system.    
		
	\end{description} 	   
	
	The first two requirements are of technical nature, stipulating that, in the randomization process, we should use ``honest dice that cannot be tampered with'' -- or else a more convenient device, like a computational algorithm that adequately mimics all the relevant characteristics of ``honest dice''. 
	For more technical discussions on theses characteristics, see Marcondes et al (2019), Saa and Stern (2019) and Silva et al. (2020).    
	
	In order to emphasize the importance of the last two requirements, let us discuss a form of cheating known as \textit{rerandomization}. 
	In this kind of cheating, the agent responsible for a sortition has the privilege of using the randomization mechanism out of public scrutiny, examine the outcome, and chose at will either to make this process and its outcome public, or to hide this first try and randomize a second time, as if the first try never happened.       
	Imagine for example the classic process of picking a ball from a transparent urn. 
	However, instead of making a live presentation, the sortition ceremony is recorded for broadcasting at a later time.    
	A dishonest agent could repeat and record the process twice, and only release the recording that best fits his or her goals, as if it were the only recording ever made.  
	It should be clear that the repeated use of this subterfuge gives the agent in charge some latitude to pick and choose, biasing the final outcome according to his or her convenience.

	\subsubsection*{Authority, Transparency and Understanding}   
	
	Why is transparency even required in a randomization process? 
	Would it no be possible, or even easier, to anchor the credibility of the process on a \textit{principle of authority}? 
	If a given authority is responsible for a randomization process, doesn't  the requirement of transparency imply an implicit doubt? 
	If so, doesn't the requirement of transparency imply disrespect for the same authority?    
	
	These are basic questions in philosophy of law, that can only be answered in a context that specifies the fundamental values and goals chosen by a given society. 
	Niklas Luhmann (1985, 1989), a celebrated scholar in philosophy of law, postulates that the fundamental goal of the justice system is -- ``the congruent generalization of normative expectations''. 
	That is, the final objective of the legal system is the construction of a harmonious society, where citizens have a coherent view of what constitutes a good set of rules for social behavior (normative expectations).        
	Moreover, a legal system should provide mechanisms that stimulate citizens to conform to normative expectations and inhibit their transgression. 
	
	This conception of law requires from every citizen a well founded trust that the justice system is efficient and fair, preferably obtained by conscious understanding of laws and regulations and their forms of implementation. 
	Moreover, a justice system conceived according to such principles is weak or fragile if sustained on blind faith on ad hoc authority, but strong and resilient if sustained by a conscious, engaged, and participative community. 
	The articles of Silva et al. (2020) and Stern (2018) expand these ideas.

	\section{Modular Arithmetic and Trusted \\ Roulettes}
	\label{sec:mod+roulettes}

	In this section we discuss intuitive ideas for how to implement an honest randomization device that all interested parties can trust. 
	In following sections of this article we offer a viable technological solution to the problem of randomization that satisfies all requirements stipulated in previous sections, following the general ideas hereby discussed.

	\begin{figure}[tb] 
		\centering   
		\includegraphics[height=34mm, 
		trim= 0mm 0mm 0mm 0mm, clip]{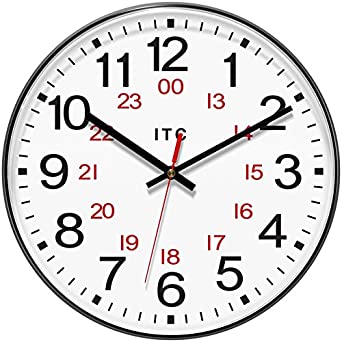} 
		\mbox{} \hspace{2mm}  
		\includegraphics[height=34mm,   
		trim= 0mm 0mm 0mm 0mm, clip]{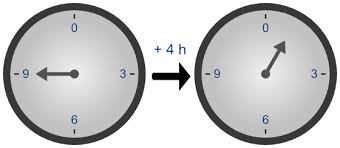} 
		
		\caption[]{Modulo 12 or Clock Arithmetic}    
		
	\end{figure}
	
	Modular arithmetic is an integer arithmetic system in which numbers ``wrap around'' after reaching a maximum value, $m$, called the modulus. 
	A familiar example is the standard reading of a clock. 
	After noon (12 o'clock), we restart counting from $1,2,...$ (p.m.).
	Notice that, in Figure 5, the position corresponding to noon (or midnight) is marked either by the modulus value, $m=12$, or by the value \textit{zero} -- that is mathematically more convenient.  
	In general, for positive integers, $n$ and $m>1$, we define $n \modulo m$ (read \textit{n modulo m}) as the remainder of the division of $n$ by $m$. 
	For example, see Figure 5 (left): $13 \modulo 12 = 1$, $14 \modulo 12 = 2$, ... $23 \modulo 12 = 11$, and $24 \modulo 12 = 12 \modulo 12 = 0$. 
	Figure 5 (right) illustrates the modular arithmetic operation $(9+4) \modulo 12 = 1$.

	Now imagine we have a game using a roulette or wheel of fortune, see Figure 5, with $k$ participants, also known as the \textit{stakeholders}, all of them wanting the privilege of spinning the roulette, and not trusting anyone else to do the job. How can we break this deadlock?

	We can solve the aforementioned impasse using the following protocol: 
	\begin{enumerate} 
		
		\item Provide each stakeholder with a  well-balanced roulette, marked according to the numbers set $\{ 0,1,2\ldots (m-1) \}$; 
		
		\item Ask each stakeholder to spin his roulette \textit{honestly}, that is, with a not fully controlled and strong enough initial impulse so to produce any of the possible outcomes,  $\{ 0,1,2\ldots (m-1) \}$,   
		with the same probability, $(1/m)$. 
		Moreover ask each stakeholder to use his roulette  
		\textit{independently}, that is, to do so without sharing any information with other stakeholders or interested parties;  
		
		\item Collect and add, using modulo $m$ arithmetic, the results produced by each one of the $k$ stakeholders in order to produce the final result: $n_f = (n_1+n_2,\ldots +n_k) \modulo m$.  
		
	\end{enumerate}

	\begin{figure}[tb] 
		\centering   
		\includegraphics[height=34mm, 
		trim= 0mm 0mm 8mm 0mm, clip]{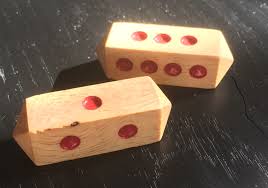} 
		\mbox{} \hspace{0mm} 
		\includegraphics[height=34mm,   
		trim= 5mm 0mm 0mm 0mm, clip]{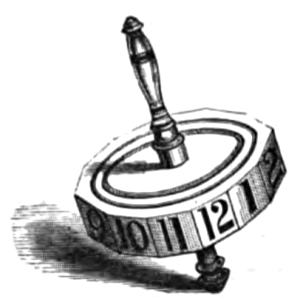} 
		\mbox{}  \hspace{2mm}    
		\includegraphics[height=34mm,   
		trim= 0mm 0mm 0mm 0mm, clip]{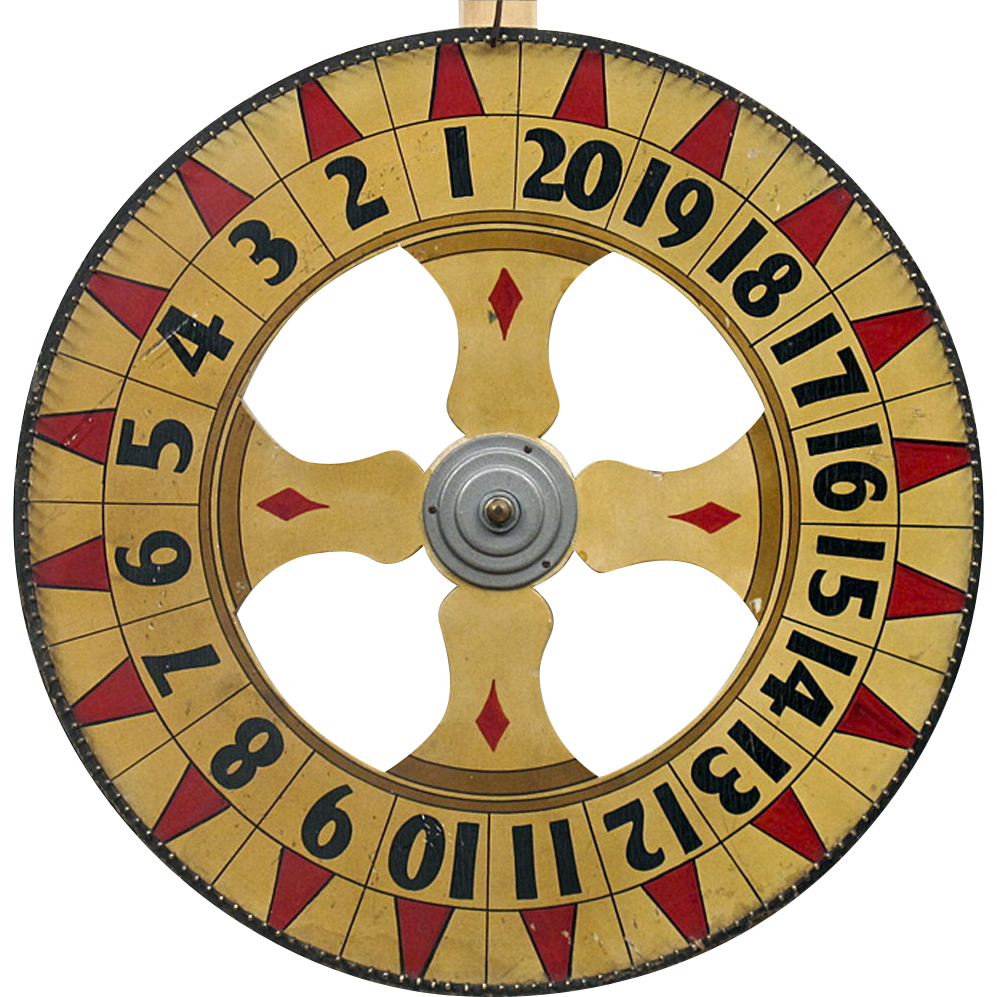} 
		
		\caption[]{Rolling Dice and Spinning Roulettes}    
		
	\end{figure}
	
	We can guarantee that the final result, $n_f$,  produced by this protocol is equivalent to an ``honest roulette'', as long as at least one (any one) of the $k$ stakeholders does his job as required. 
	This guarantee is a corollary of the following theorem: 
	Let $x$ and $y$ be independent random variables in $\{ 0,1,2\ldots (m-1) \}$. 
	Then, if any one of these random variables, $x$ or $y$, is uniformly distributed, so is $z= (x+y) \modulo m$. 
	Imagine, for example, that variable $x$ is not random at all, but rather a known constant, $c$, namely, the initial state of the roulette. 
	Furthermore, imagine that $y$ is independent of $x$ and uniformly distributed in $\{ 0,1,2\ldots (m-1) \}$. 
	Under these conditions, the theorem states that the final state of the roulette, $z=c+y$ is uniformly distributed, corresponding to the intuitive idea that, when using a well-constructed roulette, a strong enough impulse will produce a final outcome that ``forgets'' the initial state of the roulette.     
	For further details and formal mathematical analyses, see Scozzafava (1993).

	In many applications in statistics, clinical trials, and complex sortitions, we need a random variable $x$ uniformly distributed in the interval $[0,1[$ of the real line. 
	In computational procedures, this continuous variable can be approximated by a fraction $n/m$, where $m$ in a large integer, and $n \in \{0,1,2,\ldots (m-1) \}$.  
	This fraction can be translated to standard floating point notation, and then be further transformed into random variables with several probability distributions of interest in statistical modeling, see Hamersley (1964), Ripley (1987). 
	Such uniform or non-uniform random variables can, in turn, be used in dynamic clinical trials, haphazard intentional sampling, adaptive sampling procedures, and other complex applications of interest in statistical modeling and decision science, see for example Fossaluza (2015) and Lauretto et al. (2012, 2017) and the bibliography therein.    
	
	In the sequel, we describe a software implementing the protocol outlined in this section, including all necessary precautions in order to guarantee cryptographical security.   
	In this software, every stakeholder is required to input a random number $n$ between $0$ and $m=9,999,999$. 
	Each stakeholder has the responsibility of producing his or hers random 7-digit decimal number in a way he or she finds appropriate.

	\begin{figure}[tb] 
		\centering   
		\includegraphics[height=35mm, 
		trim= 18mm 5mm 22mm 4mm, clip]{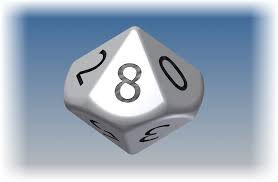} 
		\mbox{} \hspace{1mm}  
		\includegraphics[height=35mm, angle=0,   
		trim= 0mm 0mm 0mm 0mm, clip]{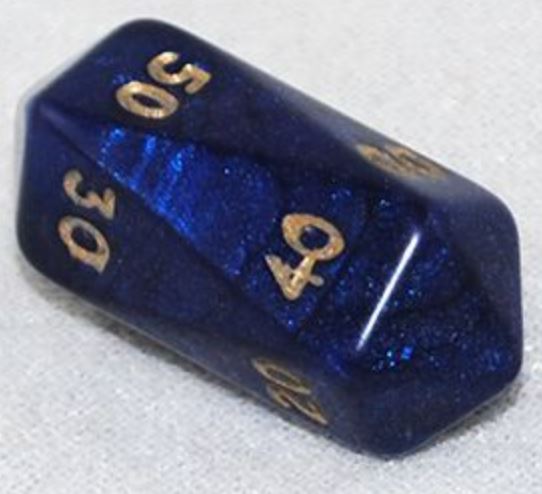} 
		\mbox{} \hspace{1mm}  
		\includegraphics[height=35mm,   
		trim= 50mm 15mm 50mm 15mm, clip]{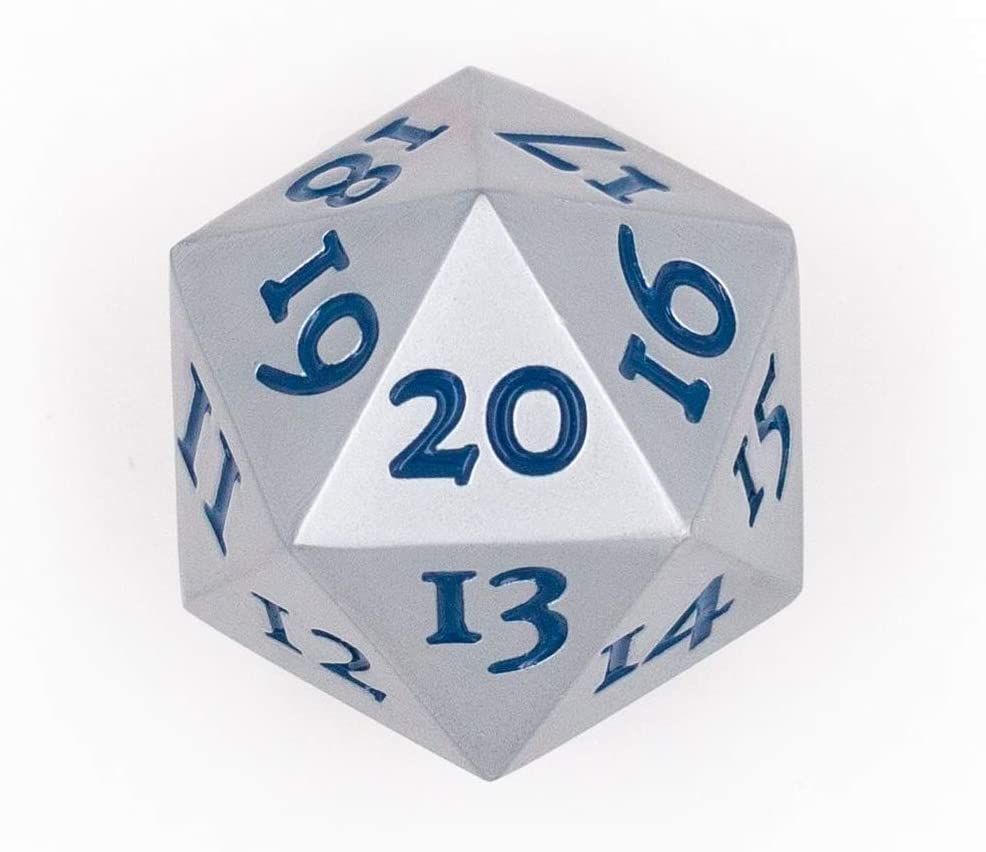} 
		
		\caption{Polyhedral dice for decimal digits}    
		
	\end{figure}  
	
	Random decimal digits can be easily produced, for example, by rolling 10-faced dice, available in several shapes with the required symmetry conditions, see Stern (2011). 
	Figure 6 exhibits 10-faced dice shaped as a pentagonal trapezohedron (left) and a pentagonal antiprism (center), and also 20-faced dice shaped as an icosahedron (right);  
	in order to produce a random decimal digit using icosahedral dice, the user should read only the last digit for outcomes ranging from 10 to 20.

	\section{Intuitive Cryptology} 
	
	
	One crucial requirement of the random drawing approach described in Section \ref{sec:mod+roulettes} is that the the roulettes are run independently.
	%
	%
	Otherwise, a dishonest stakeholder $S_i$ could wait until the results of all roulettes are revealed, and then run his/her own roulette for manipulating the final outcome of the drawing: for example, suppose that the sum of the contributions from all stakeholders except $S_i$ is $n=3$ for, say, $m= 12$; after learning this value of $n$, $S_i$ could force his/her own roulette to give $n_i=2$, thus obtaining $n_f = n + n_i \modulo m = 5$ as the final (manipulated) outcome of the drawing.
	%
	%

\begin{figure}[tb!]
		\centering
		\includegraphics[trim={17pt 16pt 78pt 18pt},clip,width=0.98\textwidth]{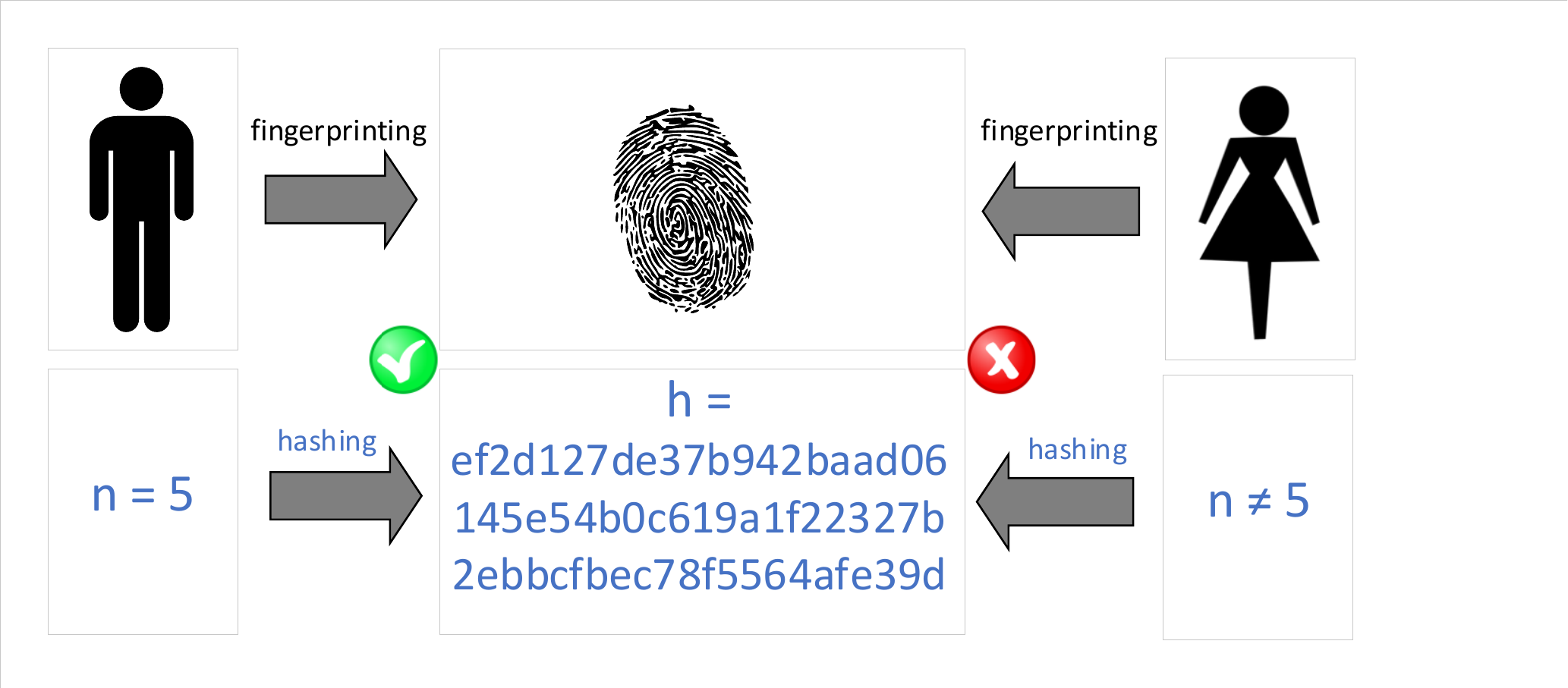}
		\caption{Hash functions and their similarity with human fingerprinting.}
		\label{fig:hash-fprint}
	\end{figure}

	To ensure this independence property, Silva et al. (2020) builds upon the properties of hash-based bit-commitment mechanisms.
	Intuitively, a hash function $H$ is a cryptography construct analogous to fingerprinting for humans, as illustrated in Figure \ref{fig:hash-fprint} -- 
	we refer the reader to Beutelspacher (1994), Bultel et al. (2017) and Fellows and Koblitz (1994) for intuitive introductions on key ideas of cryptology, an to Rogaway and Shrimpton (2004) for a concise but formal explanation of properties of cryptographic hash functions.  
	Specifically, given the fingerprint for an unknown human being, it takes a lot of computational power to look all around the world for the owner of that fingerprint; on the other hand, given a fingerprint and the corresponding human, it is quite easy to check whether or not they match, and it is hard (almost impossible) to find two different people with the same fingerprint (even considering identical twins).

	Similarly, suppose that someone computes the hash of a number $n$, i.e., a value $h = H(n)$, which acts as a ``fingerprint'' for $n$; then, if only $h$ is revealed, but $n$ is kept secret, there is no simple mechanism for finding the value of $n$.
	Of course, one could test every possible value of $n$, checking if a guess $n_i$ is such that $H(n_i) = h$, just like finding the owner of a fingerprint given only the fingerprint itself.
	However, the computational effort for performing such brute-force attack would be very large.
	Actually, in practice the cost for hash functions would be even larger than searching for a human who owns a fingerprint: while there are a few billions of humans in the world, a hash function can be used in such a manner that the number of tests would be as large as the number of atoms in the whole planet Earth!
	For this purpose, it suffices to combine the value of $n$ with a large and unpredictable (e.g., random) mask $r$ when computing the hash, i.e., to make $h = H(r, n)$.
	As a result, even if there are only a few possible values of $n$ to be tested, determining whether or not a guess $n_i$ is correct would require testing all possible values of $r$ too.
	Therefore, it suffices to use a large-enough mask $r$ (e.g., 256-bits) to ensure that any attempt of determining $n$ via brute force would be computationally infeasible.

	When both $n$ and $h$ are revealed, on the other hand, it is easy to verify whether or not they match: it suffices to compute $H(n)$ directly, and check if the output of this computation is identical to the provided value of $h$.
	However, like different humans should not have the same fingerprint, it is computationally hard to find two distinct values of $n$ (say, $n_1$ and $n_2$) that have the same hash $h$.
	Hence, once $h_1 = H(n_1)$ is revealed, one can say that the person who revealed it is ``committed'' to revealing $n_1$, i.e., it would be hard to trick someone into believing that $h_1$ was computed from any other input $n_2 \neq n_1$.
	%

	Such properties are used by Silva et al. (2020) to build a two-phase procedure for ensuring the fairness of random draws:
	
	\begin{enumerate}
		\item Commitment phase: first, each stakeholder $S_i$ runs a roulette (honestly or not), getting a value $n_i$ as result.
		Then, $S_i$ computes the hash of $n_i$, denoted $h_i$, and reveals only $h_i$ to the other stakeholders, keeping $n_i$ itself secret.
		This prevents $S_i$ from learning the roulette results from his/her peers, and vice-versa.
		
		\item Reveal phase: only after all hashes are received, every stakeholder $S_i$ reveals its own $n_i$.
		The outcome of the drawing is then computed locally by $S_i$ by adding every $n_i$ together using modular arithmetic as explained in Section \ref{sec:mod+roulettes}.
		In this case, even if $S_i$ is malicious and tries to delay the revelation of $n_i$ until it learns the partial outcome of the drawing from the values revealed by his/her peers, it would be already too late: after revealing $h_i$ in the commitment phase, $S_i$ has no choice but to reveal the already chosen $n_i$, rather than some other value that might lead to a more desirable (but unfair) drawing outcome.
		
	\end{enumerate}

	
	\section{\textit{Java} Implementation}
	
\begin{figure*}[tb!]
		\begin{subfigure}[t]{.495\textwidth}
			\centering
			\includegraphics[width=\linewidth]{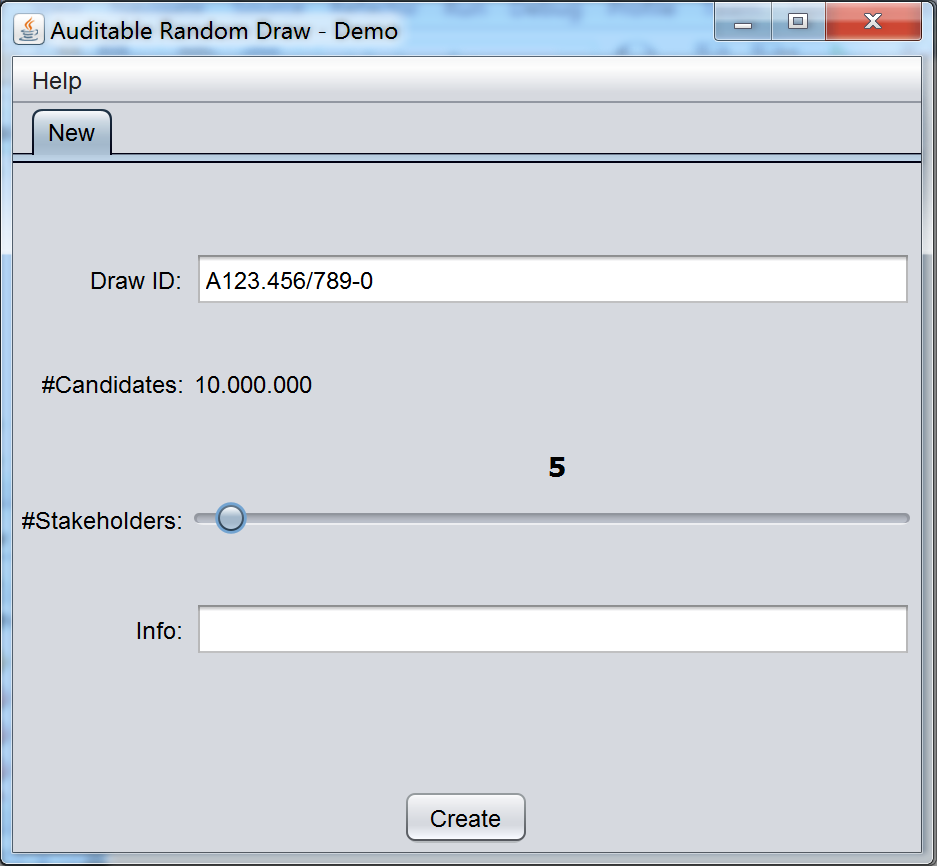}
			\caption{Random drawing with 5 stakeholders ($S_0$, $S_1$, $S_2$, $S_3$ and $S_4$) and modulus value $m=10,000,000$.}
		\end{subfigure}
		\hfill
		\begin{subfigure}[t]{.495\textwidth}
			\centering
			\includegraphics[width=\linewidth]{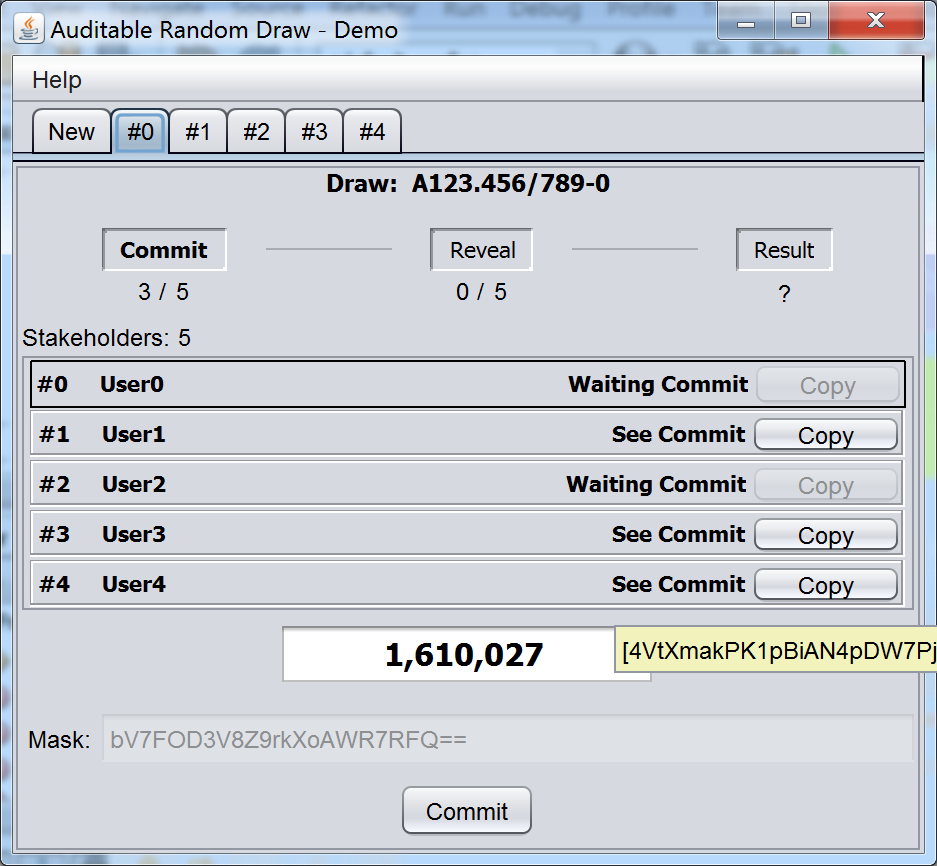}
			\caption{$S_1$, $S_3$ and $S_4$ after commitment, as seen by $S_0$. Value committed by $S_4$ is $h_4 = \mathtt{4VtCmakPK1pBiAN4pDW7Pj}\ldots$}
		\end{subfigure}
		\medskip
		\begin{subfigure}[t]{.495\textwidth}
			\centering
			\includegraphics[width=\linewidth]{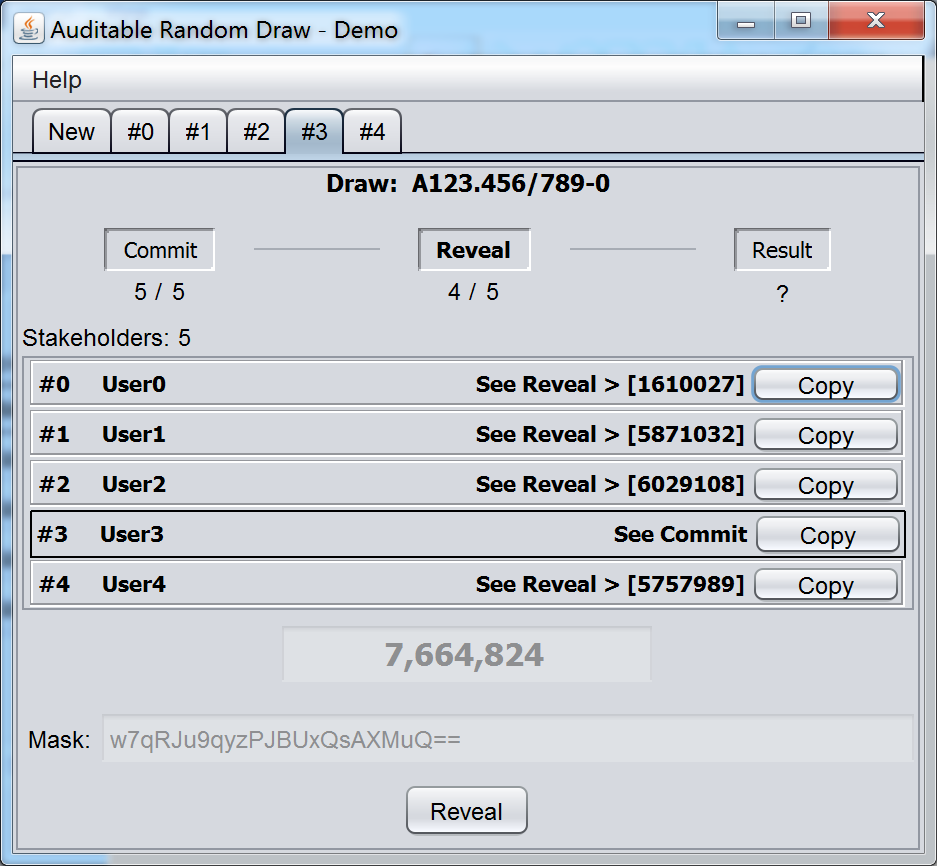}
			\caption{$S_0$, $S_1$, $S_2$ and $S_4$ in reveal phase. Partial result as seen by $S_3$.}
		\end{subfigure}
		\hfill
		\begin{subfigure}[t]{.495\textwidth}
			\centering
			\includegraphics[width=\linewidth]{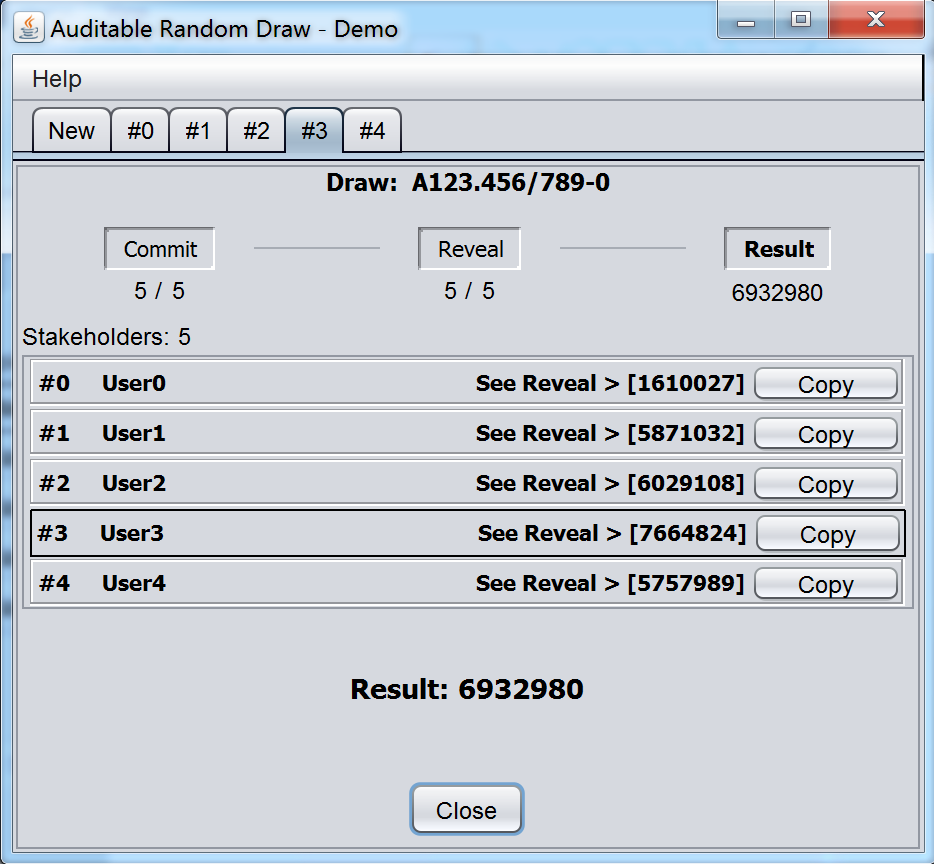}
			\caption{Drawing result: {1,610,027 + 5,871,032 + 6,029,108 + 7,664,824 + 5,757,989 {\modulo} 10,000,000 = 6,932,980.}}
		\end{subfigure}
		\caption{Proof-of-concept Java implementation: screenshots}
		\label{fig:poc+screen+simplest}
	\end{figure*}

	We developed a simple Java library for implementing the protocol described in Silva et. al (2020), and made it available under the MIT License at \url{https://doi.org/10.24433/CO.6108166.v1}
	This library can, thus, be freely adapted for the needs of specific application scenario.
	To help in this task, we also provide a simple proof-of-concept graphical interface for testing purposes, which is depicted in Figure \ref{fig:poc+screen+simplest}.
	More precisely, this figure shows:
	
	\begin{enumerate}[label=(\alph*)]
		\item A simple configuration interface for drawing a number among $m=$    10,000,000 candidates, i.e., from 0 to 9,999,999. The number of stakeholders participating in the drawing and additional metadata related to it can also be defined.
		
		\item A snapshot of the Commitment phase, as seen by Stakeholder $S_0$ in a drawing involving 5 stakeholders. In this snapshot, $S_0$ is then free to choose a number $n_0$ to commit, which is combined with a random mask for better security against brute force attacks.
		Meanwhile, $S_1$, $S_3$ and $S_4$ have already sent the hashes $h_1$, $h_3$ and $h_4$ of their own commitments, $n_1$, $n_3$ and $n_4$, respectively; as a result, these stakeholders cannot modify the chosen values $n_1$, $n_3$ and $n_4$ anymore. 
		
		\item A snapshot of the Reveal phase, as seen by Stakeholder $S_3$.
		The figure shows that $S_3$ is the only one who has not yet revealed the chosen value for $n_3$, while all of his/her peers have already revealed $n_0$, $n_1$, $n_2$ and $n_4$.
		Nevertheless, $S_3$ can only reveal the correct $n_3$ (and corresponding mask), since the revealed value must match the committed value $h_3$.
		
		\item The completion of the protocol, when one of the eligible numbers (namely, 6,932,980) is picked with uniform probability based on all stakeholders' contributions $n_0$, $n_1$, $n_2$, $n_3$ and $n_4$.
		
	\end{enumerate}
	
	\section{Final Remarks}
	
	Previous articles of this research group have explored the need of randomization procedures in legal systems, like the random assignment (distribution) of legal cases to individual judges or courts, the sortition of jurors for a given case, etc., see Marcondes et al. (2019), Saa and Stern (2019), Silva et al. (2020). 
	Moreover, these papers provide extensive discussions on how to build honest (statistically non-biased) and cryptographically secure procedures and protocols, on the sociological and political importance of using fully transparent and auditable procedures, and on the positive effects of using procedures fully compliant with the aforementioned desiderata in the constitution of strong and autonomous legal institutions. 
	
	Finally, breaking away from vicious old habits can always be stimulated by respectful criticism, by firm encouragement, and by making available user friendly tools that facilitate the adoption of virtuous new habits without the imposition of additional difficulties beyond the already heavy load of overcoming corporate inertia. 
	This paper provides such a tool, fully compliant with all technical desiderata, user friendly, written in freely available and open source code. 
	The authors hope it will be soon put to use by Brazilian legal institutions and, if necessary, stand ready to help in this endeavor.

	\subsubsection*{Acknowledgments}

	This work was supported by:
	Ripple's University Blockchain Research Initiative; 
	CNPq (Brazilian National Council for Scientific and Technological Development -- grants PQ 307648/2018-4 and 301198/2017-9); and 
	FAPESP (S\~{a}o Paulo Research Foundation, grants CEPID-CeMEAI 2013/07375-0 and CEPID-Shell-RCGI 2014/50279-4). 
	The authors are grateful for suggestions received from participants of the Interdisciplinary Colloquium on Probability Theory, held on October 10, 2019 at IEA-USP (Institute of Advanced Studies of the University of Sap Paulo), for early conversations with Julio Adolfo Zucon Trecenti from ABJ (Brazilian Jurimetrics Association), and for the mobile interface design conceived by Giovanni A. dos Santos and Joao Paulo A. S. E. Lins.  
	The authors are grateful for the invitation of Jean-Yves Beziau, from ABF (Brazilian Academy of Philosophy), and for the effort of Jos\'{e} Ac\'{a}cio de Barros and D\'{e}cio Krause, organizers of the Festschrift celebrating Doria's 75th birthday.

	\section*{References} 
	
	\renewcommand{\baselinestretch}{0.89}
	\parskip 0.70mm 
	\begin{small} 
		
		\rr A. Beutelspacher (1994). \textit{Cryptology}. Mathematical Association of America. 
		
		\rr X. Bultel, J. Dreier, P. Lafourcade, M. More (2017). How to explain modern security concepts to your children. \textit{Cryptologia}, 41, 5, 422–447. \\  \texttt{10.1080/01611194.2016.1238422}
		
		
		\rr M. R. Fellows, N. Koblitz (1994). Kid Crypto. \textit{Congressus Numerantium}, 99, 9–41. 
		
		\rr J.M. Hammersley and D.C. Handscomb (1964). \textit{Monte Carlo Methods}. Methuen.
		
		\rr V. Fossaluza, M.S. Lauretto, C.A.B. Pereira, and J.M. Stern (2015). Combining optimization
		and randomization approaches for the design of clinical trials. \textit{Springer Proceedings in
			Mathematics and Statistics}, 118, 173–184. \   \texttt{doi:10.1007/978-3-319-12454-4 14} 
		
		\rr M.S. Lauretto, F. Nakano, C.A.B. Pereira, J.M. Stern (2012). Intentional Sampling by
		Goal Optimization with Decoupling by Stochastic Perturbation. \textit{American Institute of Physics Conference Proceedings}, 1490, 189-201. \ \texttt{doi:10.1063/1.4759603} 
		
		\rr M.S. Lauretto, R.B. Stern, K.L. Morgam, M.H. Clark, J.M. Stern (2017). Haphazard
		Intentional Allocation and Rerandomization to Improve Covariate Balance in Experiments.
		\textit{American Institute of Physics Conference Proceedings},  1853,    050003, 1–8. \  \texttt{doi:10.1063/1.4985356} 
		
		\rr N. Luhmann (1985). \textit{A Sociological Theory of Law}. Routledge.
		
		\rr N. Luhmann (1989). \textit{Ecological Communication}. The Univ. of Chicago Press.
		
		
		\rr M. Naor, M. Yung (1990). Public-key cryptosystems provably secure against chosen ciphertext attacks. \textit{Proceedings of the 22nd Annual ACM Symposium on Theory of Computing}, May 13-17, Baltimore, MD, p. 427-37. 
		
		
		\rr  D. Marcondes, C. Peixoto, and J.M. Stern (2019). Assessing randomness in case assignment:
		The case study of the Brazilian Supreme Court. \textit{Law, Probability and Risk}, 18, 2/3, 97–114.
		\ \texttt{doi:10.1093/lpr/mgz006}

		\rr J. Pearl (2009). \textit{Causality: Models, Reasoning, and Inference}. Cambridge Univ. Press. 
		
		\rr B. Ripley (1987). \textit{Stochastic Simulation}. Wiley. 
		
		\rr P. Rogaway,  T. Shrimpton (2004). Cryptographic hash-function basics: Definitions,
		implications, and separations for preimage resistance, second-preimage resistance, and
		collision resistance. 
		\textit{Lecture Notes in Computer Science}, v.3017 p.371-88.   
		
		\rr O.T. Saa, J.M. Stern (2019).  Auditable Blockchain Randomization Tool. \textit{Proceedings}, 
		33, 1,, 17.1–17.6.  \  \texttt{doi:10.3390/proceedings2019033017}   
		
		\rr P. Scozzafava (1993). Uniform distribution and sum modulo m of independent random variables. \textit{Statistics and Probability Letters}, 18, 4, 313-314.  
		
		\rr M.V.M. Silva, M.A. Simplicio, R.A.C. Pfeiffer, J.M. Stern (2020).   
		\textit{A Fair, Traceable, Auditable and Participatory Randomization
			Tool for Legal Systems}.  \  \texttt{arXiv:2006.02956}                 
		
		\rr J.M. Stern (2008). Decoupling, Sparsity, Randomization, and Objective Bayesian
		Inference. \textit{Cybernetics and Human Knowing}, 15, 2, 49-68.
		
		\rr J.M. Stern (2011). Symmetry, Invariance and Ontology in Physics and Statistics.
		\textit{Symmetry}, 3, 3, 611-635.  \  \texttt{doi:10.3390/sym3030611}
		
		\rr  J.M. Stern (2018).  Verstehen (causal/interpretative understanding), Erkl\"{a}ren (law-governed description/prediction), and Empirical Legal Studies.  \textit{JITE - Journal of Institutional and Theoretical Economics}, 174, 1, 105-114. \    \texttt{doi:10.1628/093245617X15120238641866}

	\end{small} 			
	
\end{document}